\documentclass[fleqn,usenatbib]{mnras}

\usepackage{newtxtext,newtxmath}

\usepackage[T1]{fontenc}

\DeclareRobustCommand{\VAN}[3]{#2}
\let\VANthebibliography\thebibliography
\def\thebibliography{\DeclareRobustCommand{\VAN}[3]{##3}\VANthebibliography}

\usepackage{graphicx}	
\usepackage{amsmath}	
\usepackage{float}
\usepackage{placeins}
\usepackage{natbib}

\makeatletter
\xpatchcmd\NAT@citex
 {%
  \@citea\NAT@hyper@{%
    \NAT@nmfmt{\NAT@nm}%
    \hyper@natlinkbreak{\NAT@aysep\NAT@spacechar}{\@citeb\@extra@b@citeb}%
    \NAT@date
  }%
 }
 {%
  \@citea
  \NAT@nmfmt{\NAT@nm}%
  \NAT@aysep\NAT@spacechar
  \NAT@hyper@{\NAT@date}%
 }
 {}{}
\xpatchcmd\NAT@citex
 {%
  \@citea\NAT@hyper@{%
    \NAT@nmfmt{\NAT@nm}%
    \hyper@natlinkbreak{\NAT@spacechar\NAT@@open\if*#1*\else#1\NAT@spacechar\fi}%
    {\@citeb\@extra@b@citeb}%
    \NAT@date
  }%
 }
 {
  \@citea
    \NAT@nmfmt{\NAT@nm}%
    \NAT@spacechar\NAT@@open\if*#1*\else#1\NAT@spacechar\fi
    \NAT@hyper@{\NAT@date}%
 }
 {}{}
\makeatother

\AtBeginDocument{}
\AtBeginDocument{}
\AtBeginDocument{}

\title[Galaxy discs regulate SMBH growth]{Galaxy discs regulate the growth of supermassive black holes}

\author[R.J. Roberts et al.]
{Ryan J. Roberts\thanks{Email: R.Roberts1@2018.ljmu.ac.uk},
Jonathan J. Davies,
Robert A. Crain
\\
Astrophysics Research Institute, Liverpool John Moores University, 146 Brownlow Hill, Liverpool L3 5RF, UK\\
}

\date{Accepted XXX. Received YYY; in original form ZZZ}

\pubyear{2025}

\begin{document}
\label{firstpage}
\pagerange{\pageref{firstpage}--\pageref{lastpage}}
\maketitle

\begin{abstract}
We examine the relationship between the mass of present-day central supermassive black holes (SMBHs, $M_{\rm BH}$), and the stellar mass ($M_\star$) and halo mass ($M_{200}$) of their host galaxies in the EAGLE simulation, and find that scatter about these relations correlates with both halo structure and galaxy morphology. EAGLE reproduces the observed $M_{\rm BH}-M_\star$ relation, including (qualitatively) its dependence on morphology: at fixed $M_\star$, disc-dominated galaxies host less massive SMBHs than ellipticals. We show that $M_{\rm BH}$ correlates with $M_{200}$, as expected if SMBHs are regulated by processes acting on the scale of the host dark matter halo, but exhibits a tighter correlation with the halo binding energy ($E_{\rm bind}$), signalling that this quantity, which encodes information about both halo mass and halo structure, is more fundamental to $M_{\rm BH}$. As with $M_{\rm BH}-M_\star$, scatter about the $M_{\rm BH}-E_{\rm bind}$ relation is strongly correlated with morphology. Gas in the central few parsecs of galaxies with present-day discs retains strong rotational support as the galaxy grows, inhibiting inward transport and precluding periods of rapid SMBH growth by gas accretion. In galaxies destined to be present-day ellipticals, however, this rotational support is disrupted, enabling gas to be funnelled onto the central SMBH, triggering rapid growth. Evolution of the mass fraction of stars formed ex-situ indicates that this disruption is caused by galaxy-galaxy interactions and mergers. Our findings corroborate the conclusion of recent studies, based on controlled simulations of an $\sim L^\star$ galaxy, that prolonged secular galaxy evolution inhibits central SMBH growth. 

\end{abstract}

\FloatBarrier



\section{Introduction}
\label{section:intro}

Supermassive black holes (SMBHs) are found at the centres of essentially all massive galaxies \citep[e.g.][]{Kormendy-Richstone1995, Kormendy-Ho2013}. The masses of SMBHs and the properties of their host galaxy are tightly correlated \citep{Magorrian1998,Ferrarese-Merrit2000, Gebhardt2000, Tremaine2002, Marconi-Hunt2003, Haring-Rix2004,Graham-Sahu2023}, a connection that appears to extend back over a significant fraction of cosmic history \citep{Mountrichas-Buat2023}. The physics governing this relationship remains poorly understood, a particularly curious puzzle in light of the vast contrast in spatial scales between SMBH event horizons and the optical sizes of galaxies. 

Nevertheless, the correlation of SMBH mass with its host galaxy's stellar mass (or that of the bulge component), or velocity dispersion, is widely \citep[but not exclusively, see e.g.][]{JahnkeMaccio2011} interpreted as evidence that energetic feedback from SMBHs acts to simultaneously regulate their own growth, and that of their host galaxy \citep[e.g.][]{Silk-Rees1998}, by limiting the rate at which gas accretes onto the galaxy \citep[or even the galaxy's dark matter halo, e.g.][]{vandeVoort2011,Wright2024}, and subsequently onto the SMBH itself. The scenario where SMBHs are seeded in dark matter haloes, grow by the accretion of gas \citep{Soltan1982} and mergers with other SMBHs, and couple a fraction of the accreted rest-mass energy to their surroundings as feedback (usually injected as thermal or kinetic energy) is now a standard component of galaxy formation models \citep[e.g.][]{Granato2004,Springel2005,Hopkins2006,Booth-Schaye2009}. Simulations incorporating this mechanism, with suitable calibration of their poorly-constrained parameters, have been shown to yield scaling relations connecting SMBH and host galaxy properties of the observed form, to regulate and quench star formation in massive galaxies to the observed level, and to broadly reproduce the observed evolution of the cosmic star formation rate and SMBH mass densities \citep[e.g.][for a recent review see \citealt{CrainvandeVoort2023}]{Bower2006,Croton2006,Schaye2015,Henriques2015,Henden2018,Weinberger2018,Dave2019,Schaye2025}.

The inferred influence of SMBHs on spatial scales much larger than that over which they dominate the gravitational potential implies that processes acting on the scales of galaxies or their host dark matter haloes \citep{Ferrarese2002,Wyithe-Loeb2003} influence their growth. Using cosmological simulations of the co-evolution of the SMBH and galaxy populations from the OWLS project \citep{Schaye2010}, \citet{Booth-Schaye2010} demonstrated that the masses of simulated SMBHs correlate primarily with the mass of their parent dark matter halo, with a secondary dependence on halo concentration. This can be understood as a consequence of SMBHs regulating their own growth; SMBHs must grow more and inject more AGN feedback energy to expel baryons from the deeper potential wells of more massive haloes, and must inject more energy still if the halo is more concentrated. They hence concluded that the SMBH mass is fundamentally governed by the halo binding energy, $E_{\rm bind}$, and that observed correlation between SMBH mass, $M_{\rm BH}$ and host galaxy (stellar) mass, $M_\star$, is a consequence of more fundamental correlations of both $M_{\rm BH}$ and $M_\star$ with $E_{\rm bind}$.

In addition to the influence of dark matter halo properties, the importance of gas flows onto SMBHs driven by galaxy interactions (e.g. mergers) is a long-standing prediction of numerical models \citep[e.g.][]{Mihos-Hernquist1994,DiMatteo2005, Robertson2006, Hopkins2010, Dubois2015, Pontzen2017, McAlpine2020, B-M23, B-M24, B-M25}, corroborated by a growing body of observational evidence indicating that galaxy-galaxy mergers trigger AGN feedback \citep[e.g.][]{Ellison2022, Ellison25, LaMarca2025}.

The influence of mergers and halo binding energy on SMBH growth histories are challenging to separate, as (at fixed mass) haloes that have undergone more mergers exhibit a greater concentration \citep{Rey2019}, and therefore a greater central binding energy. Using zoomed cosmological simulations of an individual halo, starting from `genetically modified' initial conditions \citep{Roth2016} that enable control of galaxy merger histories, \citet{Davies2022} demonstrated that, for a fixed halo assembly time and present-day halo mass, an elevated merger mass ratio can promote the growth of the central SMBH by disrupting the rotational support of gas in the galaxy disc, and more effectively funnelling gas to the galaxy centre where it fuels SMBH growth. In a follow-up study, \citet{Davies2024} further demonstrated that in the absence of significant mergers, the formation time \citep[or, equivalently, the concentration, e.g.][]{Neto2007} of an $\sim L^\star$ galaxy's dark matter halo has little influence on the growth history of the central SMBH. They attributed this result to the presence of a strong disc in a purely secularly-evolving galaxy: in this scenario, gas that cools from the halo does not flow directly to the galaxy centre, but settles into a rotationally-supported disc, inhibiting accretion onto the SMBH. They concluded that the presence of a disc makes the growth of the SMBH less sensitive to the properties of the halo, and that disruption of the rotational support provided by the disc could be crucial for driving diversity in SMBH masses at fixed halo mass.

A corollary of the findings of \citet{Davies2022,Davies2024} is that the scatter about the $M_{\rm BH}-E_{\rm bind}$ scaling relation should correlate with galaxy morphology: if mergers aid SMBH growth, one should expect the SMBHs of present-day galaxies dominated by a rotationally-supported disc to be less massive than those at the centres of elliptical counterparts hosted by haloes with similar binding energy. 

We therefore examine the influence of galaxy morphology on SMBH mass in the EAGLE (Evolution and Assembly of GaLaxies and their Environments) simulations \citep{Schaye2015,Crain2015}. EAGLE is well suited to our purposes, as the simulations have been shown to reproduce the present-day galaxy stellar mass function, the $M_{\rm BH} - M_\star$ relation, and the demographics of galaxy morphologies \citep{Correa2017}. Moreover, a correlation between SMBH mass and galaxy morphology at fixed stellar mass in EAGLE galaxies has been shown by \citet{Correa-Schaye2020}. We extend their work to examine the connection between SMBH mass and galaxy morphology at fixed halo mass and fixed binding energy. Our paper is structured as follows. Section 2 provides a brief description of the EAGLE simulation and describes our methods to characterise galaxy and halo properties. We present our results in Section 3, focussing in particular on SMBH scaling relations, and factors governing their scatter. We summarise our findings in Section 4.
\FloatBarrier

\section{Methods}
\label{section:methods}

We examine the properties of a present-day population of simulated galaxies from the flagship EAGLE Ref-L0100N1504 simulation, drawing data from both the public database described by \citet{McAlpine2016} and directly from particle data in the simulation snapshots. For a detailed description of the EAGLE simulations and the adopted galaxy formation model, we refer readers to the project's reference articles \citep{Schaye2015,Crain2015}. Briefly, the simulation evolves a periodic volume of side $L=100\,{\rm cMpc}$, with baryonic particles of mass $\sim 10^{6}\,{\rm M_{\odot}}$, dark matter particles of mass $\sim  10^{7}\,{\rm M_{\odot}}$, and a gravitational softening length limited to $\sim 1 {\rm pkpc}$, thus resolving galaxies of mass $M_{\star} \approx 10^{9.5}\,{\rm M_{\odot}}$ with $3000$ particles. At the present day there are $\approx 325,000$ galaxies of at least this mass. 

BHs are seeded at the centres of haloes with mass $\geq 10^{10}{\rm M}_{\odot}/h$ that do not already have a BH, per \citet{Springel2005}, and accrete gas at a rate specified by a modified version of the spherically-symmetric \citet{Bondi-Hoyle1944} formula \citep[see][]{Rosas-Guevara2015}. AGN feedback is implemented as stochastic, isotropic heating of gas particles neighboring the SMBH, with a temperature increment of $\Delta T = 10^{8.5}\,{\rm K}$. \citet{Schaye2015} demonstrate that the median $M_{\rm BH}$ at fixed $M_\star$ of present-day galaxies (centrals and satellites) in EAGLE is consistent with the observational data compiled by \citet{McConnell13}, and \citet{Crain2015} show how the $M_{\rm BH}-M_\star$ relation responds to adjustment of key parameters of the EAGLE feedback model. The good correspondence is in part due to the choice of the subgrid AGN feedback coupling efficiency: this parameter was not explicitly tuned, but had the normalisation of the simulated $M_{\rm BH}-M_\star$ relation differed markedly from the observations, it would have motivated the use of a value different from the adopted $\epsilon_{\rm f}=0.15$. 

Groups and subgroups are identified using the \textsc{subfind} algorithm \citep{Springel2001,Dolag2009}, and we define galaxy-specific properties such as $M_\star$ and morphological diagnostics (see below) using the properties of the appropriate particles within $30\,{\rm pkpc}$ of the most-bound particle. We consider the SMBH mass of central galaxies to be the (sub-grid) mass of the most-massive BH particle bound to the corresponding central subhalo. We define halo mass, $M_{200}$, as the total mass contained within the virial radius $R_{200}$, the radius of a sphere with mean internal density 200 times the critical density. 

Where we examine the evolution of individual galaxies, we consider the properties of the galaxy's main progenitor halo over the redshift range $0<z<5$, where we define the main progenitor as that whose branch of the galaxy’s merger tree that contains the greatest total mass \citep[see][]{Qu2017}. We obtain the full merger tree of each galaxy from the EAGLE public database. For a simple means of illuminating when galaxy-galaxy mergers significantly influence galaxy growth, we compute the evolution of the mass fraction, $f_{\rm ex\text{-}situ}=M_{\star,{\rm ex\text{-}situ}}/M_{\star}$, of stars that formed external to the main progenitor and subsequently accreted onto it. Here, $M_{\star,{\rm ex\text{-}situ}}$ is the sum of the maximum historical stellar mass of subhaloes that merge onto the main branch of the merger tree before the next snapshot. We consider the maximum historical stellar mass in order to account for stars that may be lost from satellite galaxies due to tidal interactions with the main progenitor galaxy before the two subhaloes coalesce \citep[see e.g.][]{Qu2017, McAlpine2020}.

We quantify the morphology of simulated galaxies using three complementary diagnostics. The $\kappa_{\rm co, \star}$ parameter \citep{Correa2017} is a kinematic descriptor of a galaxy's morphology, that specifies the fraction of the stellar kinetic energy invested in motion that is co-rotational with the galaxy's bulk angular momentum,
\begin{equation}
    \kappa_{\rm{co}} = \frac{1}{K} \sum_{i, L_{z,i} > 0} \frac{1}{2} m_i
    \left(\frac{L_{z,i}}{m_i R_i^2}\right).
\end{equation}
Here $K$ is the total kinetic energy of stellar particles within a (spherical) radius of $30\,{\rm kpc}$ from the galaxy centre, $m_{i}$ is the mass of each particle,  $L_{z,i}$ is the particle angular momentum along the direction of the total angular momentum of the galaxy, and $R_{i}$ is the two-dimensional projected distance in the plane normal to the total angular momentum. By summing only over particles with $L_{z,i}>0$, we consider only the stellar particles co-rotating with the galaxy. \citet{Correa2017} show that simulated galaxies with $\kappa_{\rm co, \star} > 0.4$ exhibit colours broadly corresponding to those of the observed `blue cloud', whilst those with $\kappa_{\rm co, \star} < 0.4$ are comparable to the red sequence. We perform a similar calculation to determine the fraction of kinetic energy invested in co-rotation for all gas particles within 3kpc, which we denote as $\kappa_{\rm co,gas}$.

The $\alpha_{\rm m}$ parameter, introduced by \citet{Thob2019}, is based on the ellipticity, $\epsilon$, and triaxiality, $T$, shape parameters that are defined from the axis lengths when modelling the galaxy as an ellipsoid. The axis lengths follow from the eigenvalues of the reduced inertia tensor of the stellar mass distribution \citep[e.g.][]{Dubinski1991}. Both $\epsilon$ and $T$ can be used to describe galaxy morphology, but \citet{Thob2019} showed that neither provides a simple means of separating the blue cloud from the red sequence. They found that $\alpha_{\rm m} \equiv (\epsilon^2+1-T)/2$ cleanly separates oblate spheroids (that characterize the morphology of late-type galaxies) from spheres and prolate spheroids (characteristic of the morphology of early-type galaxies). They found that galaxies with $\alpha_{\rm m} > 0.5$ largely populate the blue cloud and those with $\alpha_{\rm m} < 0.5$ populate the red sequence. 

We also consider a simple estimate of the disc-to-total stellar mass ratio, D/T, which is assumed to be the remainder when the bulge-to-total mass fraction, B/T is subtracted from unity. The bulge is assumed to have no net angular momentum, enabling its mass to be estimated as twice the mass of stars that are counter-rotating with respect to the galaxy \citep[e.g.][]{Crain2010}. D/T is therefore defined as:
\begin{equation}
        {\rm D/T} = 1 - \frac{2}{M_{\star}}\sum_{i, L_{z,i} < 0}m_{i},
\end{equation}
where $M_{\star}$ is the total stellar mass within a spherical radius of 30kpc about the galaxy centre, $m_{i}$ is the mass of star particle $i$, $L_{z,i}$ is the particle angular momentum along the direction of the total angular momentum of the galaxy, and the sum is over all counter-rotating particles within 30kpc of the galaxy centre. \citet{Thob2019} shows that galaxies with ${\rm D/T} > 0.45$ broadly populate the blue cloud and those with ${\rm D/T} < 0.45$ the red sequence. All three diagnostics are included in the EAGLE public database.

We assign each galaxy an intrinsic binding energy, $E_{\rm bind}$, which we equate to the binding energy of the central region ($r<r_{2500}$, centred on the most-bound particle) of the halo\footnote{We have verified that using other commonly-used apertures such as $r_{200}$ and $r_{500}$ yields similar results.} in the absence of dissipative baryonic processes \citep[$E_{\rm DMO}^{2500}$ in the nomenclature of][]{Davies2019}. We elect to eliminate the influence of baryonic processes because they can dynamically alter the structure of the halo, for example via strong outflows. \citet{Davies2020} demonstrate that the central binding energy correlates strongly with the ratio $V_{\rm max}/V_{200}$, a commonly-used proxy for halo concentration, at fixed halo mass. 

We compute $E_{\rm bind}$ by summing the binding energies of all particles within $r_{2500}$ for the counterpart of each halo identified in the DM-L100N1504 simulation \citep[introduced by][]{Schaller2015}, which starts from the same initial conditions as Ref-L100N1504, but adopts purely collisionless dynamics. The counterpart haloes are identified using the bijective particle matching algorithm described by \citet{Schaller2015}. We adopt the convention that binding energy is a positive quantity, i.e. the energy required to unbind the halo, such that a larger value corresponds to a more tightly-bound halo. Particle binding energies are computed by \textsc{subfind}, with their gravitational potentials measured under the assumption that the halo is isolated and its boundary represents the potential zero point, and their kinetic energies calculated in the centre-of-mass frame of the halo's central subhalo. The halo boundary and, by extension, the centre-of-mass frame, are computed using an iterative unbinding procedure. Since the potential zero point is fixed to the halo boundary, the measured central binding energy can in principle be influenced by the halo's outer structure or local environment. However, as we consider the binding energy only of central galaxies, for which the halo boundary is unlikely to be influenced by neighbouring structures, we do not expect this effect to be significant. Where we use halo concentrations, these are obtained from the best-fit Navarro-Frenk-White profile \citep{NFW1996,NFW1997} for the radial density profile of the counterpart haloes.

Unless stated otherwise we consider a sample comprised of the 3,401 central galaxies (i.e. those hosted by central subhaloes) that have present-day mass $M_{200} > 10^{11.5}\,{\rm M_{\odot}}$ and that have an identified counterpart halo in the DM-L100N1504 simulation. We consider only central galaxies as halo mass and halo binding energy are ill-defined quantities for satellite galaxies. 
\FloatBarrier

\section{Results}
\label{section:results}

\subsection{SMBH scaling relations}
\label{section:results-scaling-relations}

\begin{figure}
            \includegraphics[width=1\linewidth]{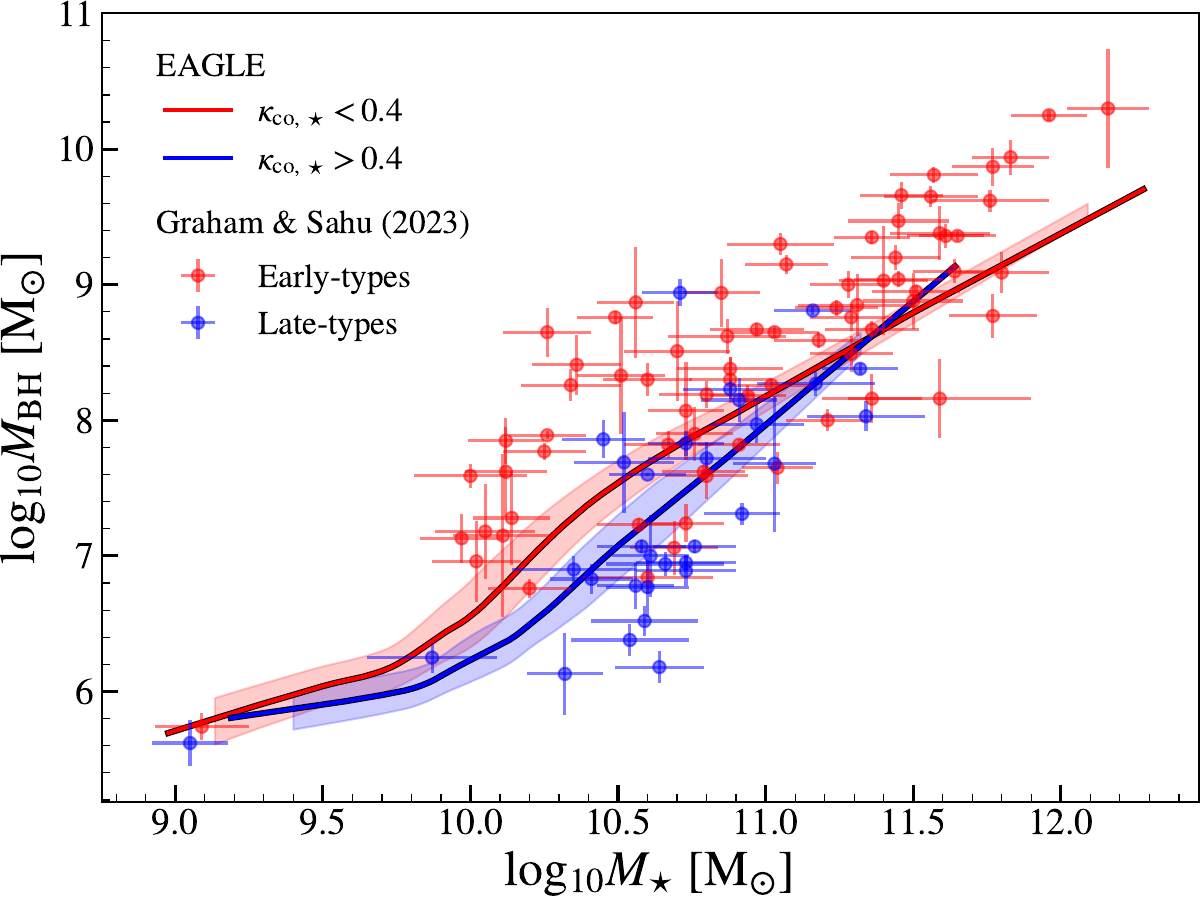}
            \caption{The running median present-day SMBH mass, as a function of galaxy stellar mass, for all early-type (red curve) and late-type (blue curve) simulated galaxies with $M_\star > 10^{9.5}\,{\rm M}_\odot$, differentiated using a threshold of $\kappa_{\rm co, \star}=0.4$ (see text for further details). Shaded regions correspond to the interquartile range. Symbols with error bars denote the compilation of observational measurements by \citet{Graham-Sahu2023}, with red and blue symbols corresponding to early- and late-type galaxies, respectively. The agreement between the simulations and the median of the observational data is good, and the simulations exhibit a qualitatively similar trend to the observations with more massive SMBHs hosted by early-type galaxies at fixed stellar mass. However the simulations exhibit less scatter, and a milder difference between early- and late-type galaxies, than is observed.}
            \label{fig:mbh_mstar}
\end{figure}

We begin by establishing that EAGLE's median present-day $M_{\rm BH}-M_\star$ relation is consistent with state-of-the-art observational measurements, and that the relation exhibits a qualitatively similar trend with galaxy morphology. Fig.~\ref{fig:mbh_mstar} shows the median $M_{\rm BH}$ as a function of $M_\star$ for present-day EAGLE galaxies with $M_\star > 10^{9.5}\,{\rm M}_\odot$ split into disc-dominated ($\kappa_{\rm co, \star} > 0.4$, blue curve) and disc-poor ($\kappa_{\rm co, \star} < 0.4$, red curve) sub-samples. In each case the shaded region corresponds to the interquartile range (IQR) of $M_{\rm BH}$ at fixed $M_\star$. We include satellite galaxies here, as the observational data we compare to does not distinguish centrals and satellites. Overlaid data points correspond to the dynamical mass measurements of SMBHs hosted by nearby galaxies presented by \citet{Graham-Sahu2023}, with filled red points corresponding to early-type galaxies (elliptical, ellicular and lenticular types) and blue points corresponding to late-type galaxies. The median relations of the two simulation sub-sets remains compatible with this recent compilation but, as was noted by \citet{Schaye2015}, EAGLE yields less scatter about the relation than observed, which is a common shortcoming of the current generation of galaxy formation simulations \citep{Habouzit2021}. 

The figure highlights that EAGLE exhibits the same qualitative trend with morphology exhibited by the observations: at fixed $M_\star$, SMBHs hosted by early-type galaxies are more massive than those hosted by late-type galaxies. At $M_\star = 10^{10.5}$ ${\rm M}_\odot$, the median SMBH mass of early-type and late-type galaxies in EAGLE is, respectively, $10^{7.58}$ $\rm M_{\rm \odot}$ and $10^{7.08}$ $\rm M_{\rm \odot}$. This result is consistent with the findings of \citet{Correa-Schaye2020}, who show that the residual SMBH masses at fixed $M_{\star}$ of EAGLE galaxies with $M_{\star} > 10^{10}$ $\rm{M}_{\odot}$ are strongly anti-correlated with residual $\kappa_{\rm co}$. Despite the simulation exhibiting less scatter than inferred from observations, the offset of the median relations for early- and late-type galaxies with mass $M_\star \approx 10^{10.5}$ ${\rm M}_\odot$ is greater than the IQR about either of the populations. A correlation between SMBH mass and galaxy morphology appears to be a common prediction of the current generation of cosmological simulations. \citet{Li2020} analysed the IllustrisTNG \citep{Pillepich2018,Nelson2018} simulations, and found that at fixed galaxy velocity dispersion, the most massive SMBHs tend to be hosted by galaxies with a high Sérsic index. \citet{Smethurst2024} analysed the Horizon-AGN \citep{Dubois2014} simulations, and found that disc-dominated galaxies have a median $M_{\rm BH}$ that is lower than the median of the overall population at fixed stellar mass, and significantly lower than that of galaxies with rich merger histories (typically associated with present-day ellipticals).

\begin{figure}
    \centering
    \includegraphics[width=\columnwidth]{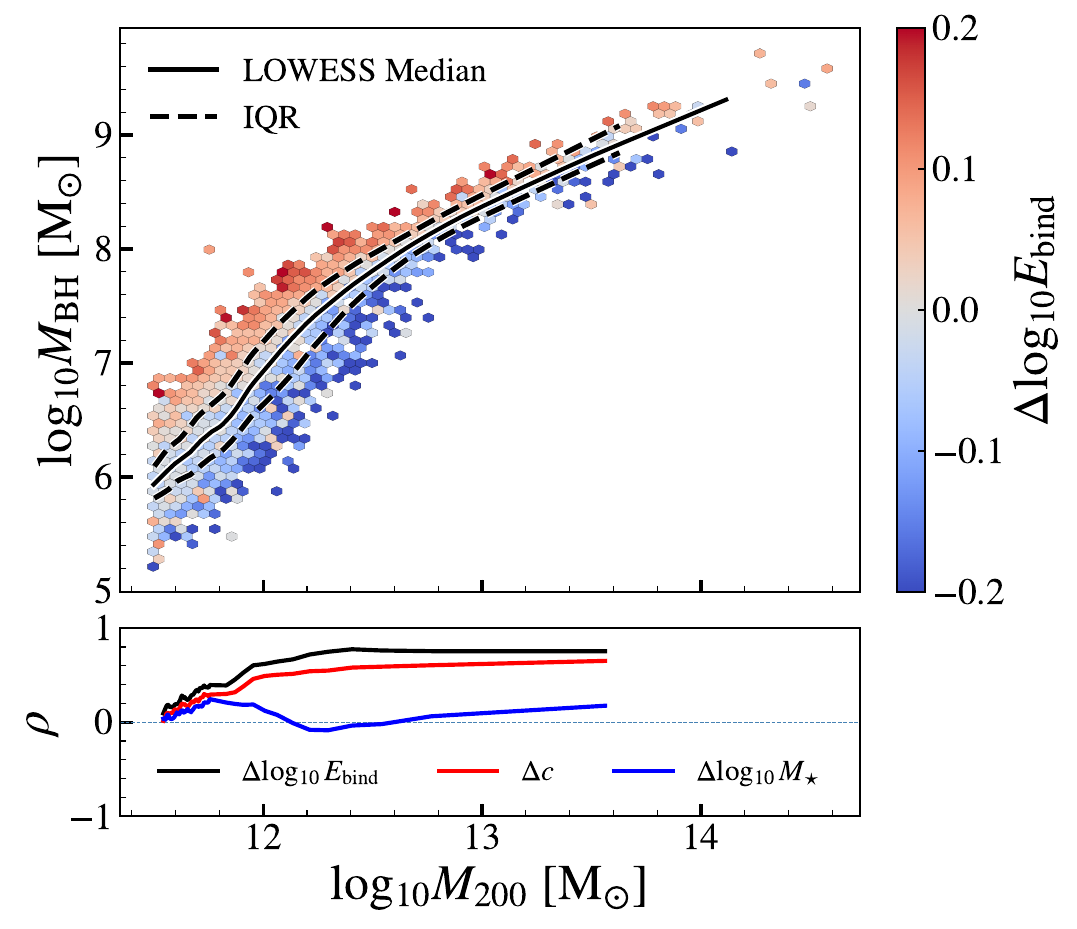}
    \caption{The median (solid curve) and IQR (dashed curves) of the SMBH mass of present-day central galaxies in EAGLE as a function of their halo mass. The curves are overlaid on hexbins whose colour encodes the mean value of $\Delta \log_{10}\,E_{\rm bind}$, the logarithmic residual of each halo's binding energy with respect to the median halo binding energy as a function of $M_{200}$. The colouring highlights the strong correlation, at fixed halo mass, between SMBH mass and the halo binding energy, over effectively the entire range of halo masses in our sample. The strength of this correlation as a function of halo mass is quantified in the sub-panel, which shows the value of the Spearman rank correlation coefficient computed for $\Delta \log_{10}E_{\rm bind}$ (black), $\Delta c$ (red), and $\Delta \log_{10}M_{\star}$ (blue), respectively. The strongest correlation is exhibited by $\log_{10}\,E_{\rm bind}$.}
    \label{fig:mbh_m200}
\end{figure}
\mbox{}

We turn next to the influence of halo structure on $M_{\rm BH}$. Fig.~\ref{fig:mbh_m200} shows the median $M_{\rm BH}$ at fixed $M_{200}$ with a solid black curve, and the two dashed black curves denote the IQR. We compute the median $M_{\rm BH}$ and the second-order scatter correlations between $M_{\rm BH}$ and other galaxy and halo properties at fixed $M_{200}$ using locally-weighted scatterplot smoothing \citep[LOWESS,][]{Cleveland1979}. This obviates the need to bin galaxies and yields a formal value of the median evaluated at the halo mass of every galaxy in the sample. $M_{\rm BH}$ scales strongly with $M_{200}$, and a gradual change of slope in the relation is visible as BHs transition from a rapidly-growing phase in lower-mass haloes to a self-regulated growth phase in higher-mass haloes \citep[e.g.][]{McAlpine2018}. There is significant scatter at fixed halo mass: at $M_{200}=10^{12.3}\,{\rm M}_\odot$ the two middle quartiles span 0.44 dex in $M_{\rm BH}$. We show that the scatter about the median relation correlates with halo structure by overlaying the curves on hexagonal bins whose colour encodes the mean value of $\Delta \log_{10}\,E_{\rm bind}$, the logarithmic residual of each halo's binding energy with respect to the median $E_{\rm bind}-M_{200}$ relation, again computed using LOWESS. We colour by this residual rather than the binding energy itself since the latter is a strong function of halo mass. 

A strong correspondence (at fixed $M_{200}$) between $M_{\rm BH}$ and $E_{\rm bind}$ is immediately apparent: more massive SMBHs are hosted by haloes with a greater binding energy, as found by \citet{Booth-Schaye2010} in the OWLS simulations. We quantify the strength of the correlation between $\Delta \log_{10} M_{\rm BH}$ and $\Delta \log_{10} E_{\rm bind}$ by plotting in the lower panel as a black curve the running Spearman rank correlation coefficient of the two quantities. A strong correlation is seen for $M_{200}\gtrsim 10^{12}\,{\rm M}_\odot$, with a peak value of $\rho_{\rm max}=0.78$. The red and blue curves show the running Spearman coefficient of the correlations between $\Delta \log_{10} M_{\rm BH}$ and, respectively, $\Delta c$ and $\Delta \log_{10} M_\star$. The former exhibits similar behaviour to the $\Delta \log_{10} E_{\rm bind}$ case, as expected since at fixed halo mass the binding energy and concentration are strongly correlated. The peak Spearman coefficient in this case is $\rho_{\rm max}=0.65$. The correlation with stellar mass is weak for all $M_{200}\gtrsim 10^{12}\,{\rm M}_\odot$, indicating that stellar mass has little influence on the relation between the mass of an SMBH and that of its host halo. The moderate correlation seen for $M_{200} \approx 10^{11.75}\,{\rm M}_\odot$ likely follows from the correlation between $M_\star$ and $c$ at fixed $M_{200}$ in EAGLE \citep{Matthee2017}. Our analysis of EAGLE therefore indicates that, as concluded by \citet{Booth-Schaye2010} from their analysis of OWLS, SMBHs in EAGLE self-regulate as a result of processes acting on the scale of galaxy haloes, and their mass follows primarily from the halo binding energy. 

We note briefly that if one instead examines the influence of halo mass on BH mass at fixed binding energy, there is no significant correlation at low binding energy ($\log_{10} E_{\rm bind} [\rm erg] < 58$), but at higher $E_{\rm bind}$ an anti-correlation ($\rho_{\rm max} = -0.45$) emerges, which can be attributed to the halo merger history. Significant mergers can increase the halo concentration \citep{Rey2019}, therefore a particular binding energy can be indicative of a relatively massive halo that assembled secularly, or a less massive halo that experienced a greater number of significant mergers. Since mergers can also enhance BH growth \citep[e.g.][]{Davies2022}, lower mass haloes typically host more massive BHs at fixed binding energy, and vice versa. Similarly, at fixed binding energy there is a strong anti-correlation between stellar mass and BH mass for haloes with $58 \lesssim \log_{10} E_{\rm bind} [\rm erg] \lesssim 59$, which stems from stellar mass being more sensitive to halo mass than halo concentration.

\subsection{The influence of galaxy morphology}
\label{section:results-morphokinem}
\begin{figure}
    \centering
    \includegraphics[width=\columnwidth]{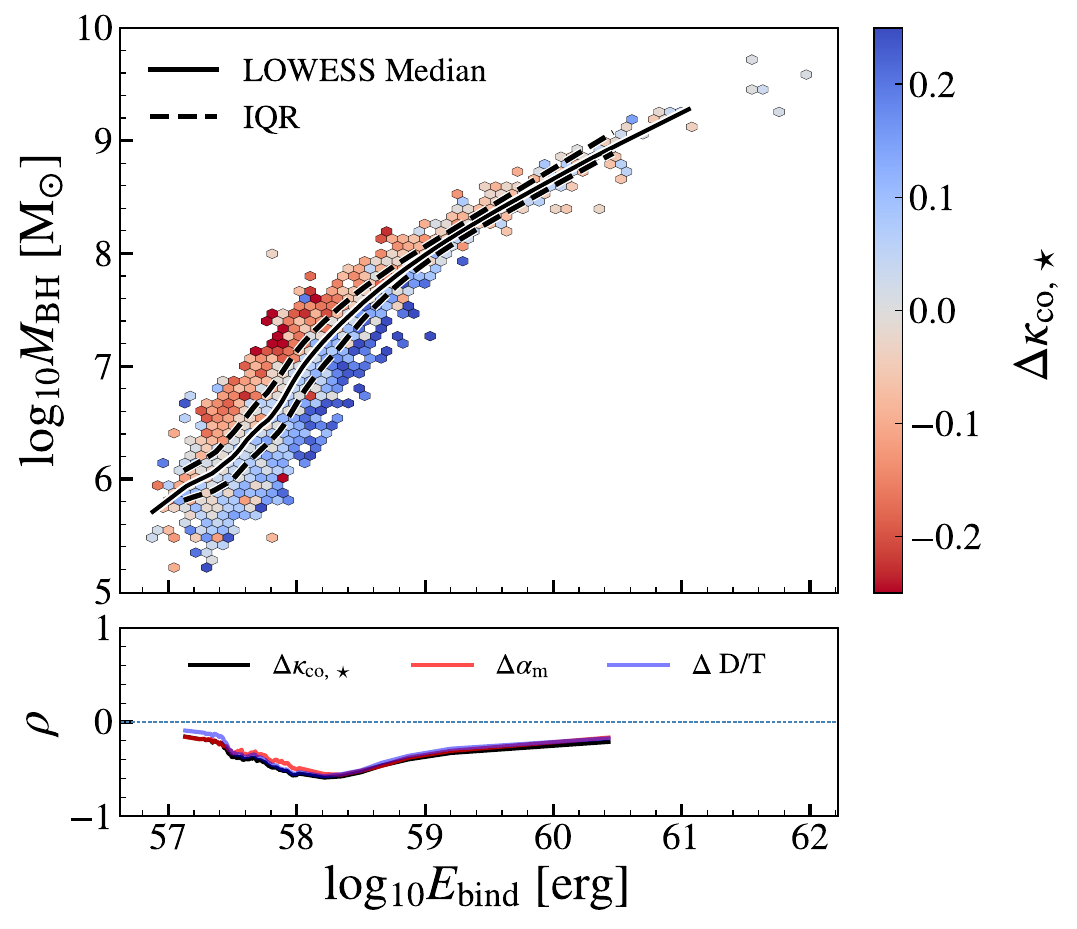}
    \caption{The running median (solid curve) and IQR (dashed curves) of the SMBH mass of present-day central galaxies in EAGLE as a function of their halo binding energy. The curves are overlaid on hexbins whose colour encodes the mean value of the residual of $\kappa_{\rm co, \star}$ with respect to its running median as a function of the halo binding energy. The colouring highlights the strong anti-correlation, at fixed halo binding energy, between SMBH mass and $\kappa_{\rm co}$ (which quantifies the `disciness' of the galaxy), over effectively the entire range of binding energies in our sample. As in Fig.\ref{fig:mbh_m200}, the sub-panel shows the running Spearman rank coefficient of the correlation between SMBH mass and $\kappa_{\rm co, \star}$ (black), $\alpha_{\rm m}$ (red) and D/T (blue). The three morphology diagnostics yield near identical correlations, each demonstrating that galaxy morphology accounts for the majority of the scatter in SMBH mass at fixed binding energy.}
    \label{fig:mbh_ebind}
\end{figure}

Having established the influence of the halo binding energy on the BH mass, we now examine the additional role played by galaxy morphology at fixed $E_{\rm bind}$. Fig.~\ref{fig:mbh_ebind} shows the relation between $M_{\rm BH}$ and $E_{\rm bind}$, with the solid black line denoting the median $M_{\rm BH}$ at fixed $E_{\rm bind}$ (computed using LOWESS) and the dashed lines corresponding to the IQR. The hexbins are coloured according to the mean enclosed values of $\Delta \kappa_{\rm co, \star}$. The sub-panel shows the running Spearman rank coefficient of the correlation between $\Delta \log_{10} M_{\rm BH}$ and $\kappa_{\rm co, \star}$ (black), $\alpha_{\rm m}$ (red) and D/T (blue), respectively. SMBH mass increases steeply as a function of halo binding energy, and the scatter about this relation is smaller than for the $M_{\rm BH}-M_{200}$ relation. At $\log_{10} E_{\rm bind} [{\rm erg}] = 58.3$, the median halo binding energy corresponding to the halo mass ($M_{200}=10^{12.3}\rm M_{\odot}$) at which binding energy most strongly correlates with SMBH mass, the IQR of $M_{\rm BH}$ is 0.36 dex, compared with 0.44 dex for the $M_{\rm BH}-M_{200}$ relation. These findings corroborate the conclusion of \citet{Booth-Schaye2010} that SMBH mass is primarily governed by the halo binding energy. 

Moreover we find that at fixed binding energy, scatter in $M_{\rm BH}$ is strongly anti-correlated with morphological diagnostics that quantify the `disciness' of the galaxy: at fixed binding energy, galaxies that are more (less) disc-dominated host a less (more) massive SMBH. This correlation is significant across nearly the entire range of halo binding energies in our sample, with the peak of the correlation with each morphological diagnostic ($\rho_{\rm max} = [-0.59,-0.56,-0.58]$ for $\kappa_{\rm co, \star}$, $\alpha_{\rm m}$ and D/T respectively) at $\log_{10} E_{\rm bind} [\rm erg] \sim 58-58.5$. The similarity of the correlation for each of the three morphological diagnostics is perhaps unsurprising given that \citet{Thob2019} demonstrated that they are strongly correlated with one another. Analysis of the simulated galaxy population in EAGLE therefore corroborates the physical picture inferred by \citet{Davies2022,Davies2024} using simulations of an individual galaxy, that galaxy discs inhibit the inward transport of gas and its subsequent accretion onto the central SMBH.

\subsection{The evolution of SMBH mass and galaxy morphology}
\label{section:results-timeevolution}

\begin{figure}
    \includegraphics[width=1\linewidth]{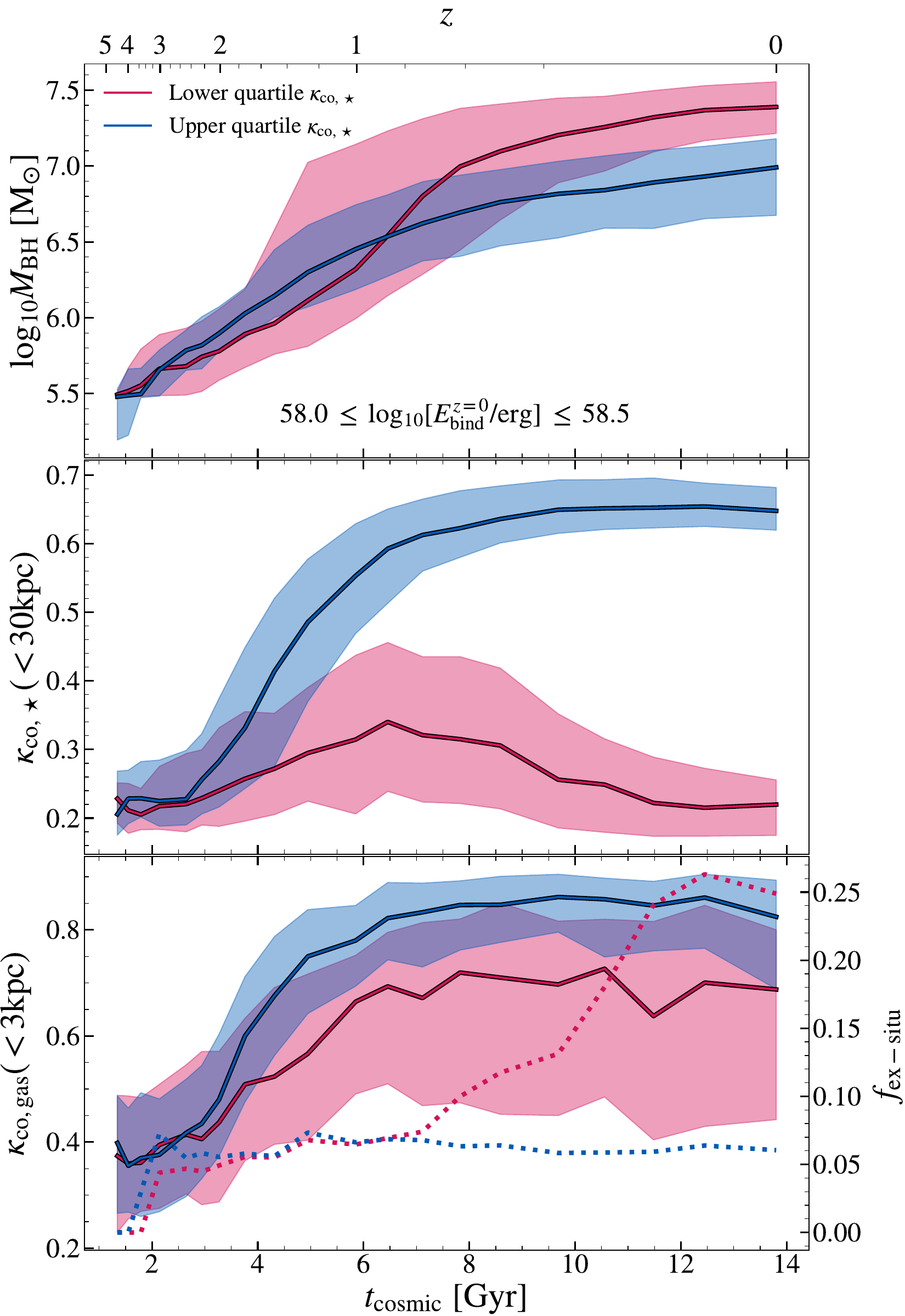}
    \caption{Evolution of the median SMBH mass ($M_{\rm BH}$, upper panel), co-rotational stellar kinetic energy fraction ($\kappa_{\rm co, \star}$, centre panel), co-rotational kinetic energy fraction for gas in the inner 3 kpc ($\kappa_{\rm co, gas}$, lower panel, solid curves, left $y$-axis), and stellar ex-situ fraction ($f_{\rm ex-situ}$, lower panel, dotted curves, right $y$-axis) for galaxies within the halo binding energy window over which the influence of morphology on SMBH mass is most apparent ($58 < \log_{10}\,E_{\rm bind} {\rm [erg]} < 58.5$). The sample is split into `most-discy' galaxies (those in the upper quartile of $\kappa_{\rm co, \star}$, blue curves) and `least-discy' galaxies (bottom quartile of $\kappa_{\rm co, \star}$, red curves). Shaded regions denote the IQR (not shown for $f_{\rm ex\text{-}situ}$, for clarity). The median SMBH mass in both subsamples evolves similarly until $t\approx 6\,{\rm Gyr}$, when the SMBHs hosted by the least discy galaxies experience a period of rapid growth that is broadly co-temporal with the decline of $\kappa_{\rm co, \star}$. The increase in ex-situ fraction following this period for these galaxies indicates that this disc disruption is caused by galaxy interactions that lead to mergers. In contrast, galaxies with the strongest present-day discs formed them early, and retained them throughout their lifetimes. The lower panel shows that the most-discy galaxies have strongly-rotating gaseous cores by present-day; for the least-discy galaxies this rotation is disrupted prior to rapid SMBH growth, after which $\kappa_{\rm co, gas}$ exhibits large scatter as there is little gas remaining due to AGN-driven expulsion.}
    
    \label{fig:mbh_history}
\end{figure}

To obtain a view of how galaxy morphology influences SMBH evolution, we identify the 573 galaxies in the narrow window of present-day halo binding energy $58 < \log_{10}\,E_{\rm bind} {\rm [erg]} < 58.5$ over which the influence of morphology is most pronounced. We then rank these galaxies by their present-day value of the co-rotational stellar kinetic energy fraction ($\kappa_{\rm co, \star}$), and select galaxies in the upper and lower quartiles to create `most discy' and `least discy' galaxy subsamples\footnote{Using instead $\alpha_{\rm m}$ or D/T rather than $\kappa_{\rm co, \star}$ does not significantly alter these samples.}, to most clearly demonstrate the impact of galaxy morphology on SMBH evolution. The upper panel of Fig.~\ref{fig:mbh_history} shows the median SMBH mass of the main progenitors of these galaxies \citep[identified using the \textsc{D-Trees} algorithm,][]{Jiang2014} as a function of cosmic time, with the most- and least-discy subsamples shown with blue and red curves, respectively. The centre panel shows the corresponding evolution of the co-rotational stellar kinetic energy fraction ($\kappa_{\rm co, \star}$) of the two subsamples. The lower panel shows the median evolution of the co-rotational kinetic energy fraction of gas in the inner 3 kpc, $\kappa_{\rm co,gas}$ (solid curves corresponding to the left $y$-axis), and the stellar ex-situ fraction $f_{\rm ex\text{-}situ}$ (dotted curves corresponding to the right $y$-axis). Shaded regions denote the IQR which, for clarity, is not shown for $f_{\rm ex\text{-}situ}$,

Prior to $t \approx 6\,{\rm Gyr}$ the subsamples exhibit similar median SMBH growth histories, evolving steadily from the seed mass to $\log_{10} M_{\rm BH} [{\rm M}_\odot] \approx 6.25$. The median SMBH mass of the least-discy galaxies then exhibits rapid growth, increasing by nearly an order of magnitude within $\approx 2\,{\rm Gyr}$, whilst that of the most-discy galaxies maintains steady growth, increasing by only $\approx 0.25\,{\rm dex}$ over the same period. Inspection of the evolution of the halo mass of the main progenitors of the two subsamples (not shown, for brevity) reveals that the onset of rapid growth in the least discy population broadly corresponds to the epoch at which the host halo reaches the `critical mass' \citep[see][in particular their eq. 9]{Bower2017} at which outflows driven by EAGLE's star formation feedback cease to be buoyant within the ambient circumgalactic medium, precluding the efficient transport of gas away from the galaxy. \citet[][see also \citealt{Dubois2015,McAlpine2018}]{Bower2017} argue that the resulting build up of gas in the halo centre enables the central SMBH to then grow rapidly until it is massive enough to self-regulate via AGN feedback. The most-discy sample exhibits a similar median halo mass growth history (in fact this sample reaches the critical mass at a slightly earlier epoch than the least-discy sample), but despite reaching the critical halo mass, these galaxies do not exhibit rapid SMBH growth, and instead continue to grow steadily. By the present-day, the median SMBH of the two populations differs by $\approx 0.4\,{\rm dex}$.

The central panel of Fig.~\ref{fig:mbh_history} highlights that the median $\kappa_{\rm co, \star}$ of the least-discy population only gradually increases until $t\approx 6.5\,{\rm Gyr}$, before exhibiting a steady decline that persists to the present-day. The median $\kappa_{\rm co, \star}$ remains at a consistently low value, though individually two-thirds of the galaxies' main progenitors can be found with high $\kappa_{\rm co, \star}$ values representative of disc-like structure in at least one snapshot, and we note that the temporal resolution afforded by the snapshots is relatively poor. The onset of the decline in the median $\kappa_{\rm co, \star}$ for this subsample is broadly co-temporal with the onset of rapid growth for the corresponding SMBHs in the upper panel. In contrast, the median $\kappa_{\rm co, \star}$ of the most-discy population increases rapidly between $t\approx 4-6\,{\rm Gyr}$ and continues rising until $t\approx 10\,{\rm Gyr}$ when it peaks at $\kappa_{\rm co, \star}\approx 0.66$ and remains at approximately this level to the present day. The survival of a stellar disc so strongly dominated by rotational support implies that this subsample of galaxies has experienced few (if any) significant mergers capable of altering the galaxy morphology \citep[e.g.][]{Stewart2008,Tacchella2019,Dillamore2022}.

The decline in stellar co-rotation for the least-discy galaxies is the result of galaxy mergers, and the tidal forces that occur as galaxies interact pre-coalescence; these events contribute stars to the dispersion-supported stellar bulge and stellar halo components of the galaxy, and disrupt the stellar disc \citep{Clauwens2018,Pfeffer2023}. The role of galaxy interactions is demonstrated by the evolution of $f_{\rm ex-situ}$ in the lower panel of Fig.~\ref{fig:mbh_history}; on average, galaxies in the least-discy sample gain approximately one quarter of their present-day stellar mass from merger events, whereas the median $f_{\rm ex-situ}$ of the most-discy sample indicates predominantly secular evolution, largely uninterrupted by mergers. The median ex-situ fraction for the least-discy galaxies increases 1-2 Gyr after their $\kappa_{\rm co, \star}$ values begin to decline and their SMBHs start to grow rapidly; merging galaxies can begin interacting long before coalescence \citep[if, for example, the infalling galaxy is on a tangential orbit rather than radially infalling, see e.g.][]{Zeng21}, and the tidal forces generated during an ongoing merger can disrupt co-rotational support long before we see any increase in $f_{\rm ex-situ}$. Mergers may therefore influence galaxy morphology - and consequently SMBH evolution - prior to the merger being dynamically complete.

The disruptive influence of galaxy interactions and mergers transfers angular momentum away from the gas in galaxies, funnelling the gas towards the central SMBH and facilitating the enhanced growth seen in our least-discy sample. We show the co-rotational kinetic energy fraction of gas in the inner 3 kpc, $\kappa_{\rm co,gas}$, in the lower panel of Fig.~\ref{fig:mbh_history}. For the most-discy galaxies, co-rotational support of the central gas increases in concert with that of the galaxy's stars (within $30 {\rm kpc}$, centre panel), with a median $\kappa_{\rm co,gas}>0.8$ for $t \gtrsim 6\,{\rm Gyr}$, indicative of galaxies evolving secularly, undisturbed by galaxy interactions. The bulk of the gas in the SMBH's vicinity is thus maintained in a co-rotating disc, and SMBH growth is inhibited, despite the fact that the host halo has grown above the `critical mass' of \citet{Bower2017}, and feedback associated with star formation can no longer efficiently expel gas from the galaxy. For the least-discy galaxies, co-rotation in the central gas is suppressed from $t\approx 4$ Gyr onwards, reflecting the overall disruption of the stellar disc on larger spatial scales. This allows more gas to reach the SMBH and fuel subsequent growth. The median $\kappa_{\rm co,gas}$ of the least-discy galaxies remains high relative to the low values seen for the galaxy-scale stellar component in the central panel, however there is significant scatter in this quantity. The enhanced SMBH growth for these galaxies yields strong AGN feedback that depletes the inner 3 kpc of gas, leaving few particles remaining from which to measure $\kappa_{\rm co,gas}$. The median and scatter shown for $t \gtrsim 6\,{\rm Gyr}$ is thus an aggregation of noisy values for most galaxies in the sample (for an example of this for an individual galaxy, see \citealt{Davies2022}, their Fig. 4b). The stellar content of the galaxies is unaffected by this feedback, and exhibits more clearly that little co-rotational motion remains in the galaxies.
\FloatBarrier

\section{Summary}
\label{section:discussion-conclusions}

We have examined the relationships between the SMBH mass, and both galaxy stellar mass and halo binding energy, and the influence on the scatter about these relations induced by galaxy morphology, using the EAGLE Ref-L100N1504 simulation. Our findings are summarised as follows:

\begin{itemize}

    \item EAGLE's present-day galaxy population exhibits a median SMBH - galaxy stellar mass scaling relation that agrees well with state-of-the-art observational measurements, and exhibits a qualitatively similar trend with morphology, such that at fixed stellar mass, early-type galaxies host more massive SMBHs than late-type galaxies. However, EAGLE exhibits less scatter in SMBH mass at fixed stellar mass than observed, and the influence of morphology on SMBH mass is also weaker than observed (\autoref{fig:mbh_mstar}).
    
    \item The SMBH mass of central galaxies in EAGLE correlates strongly with halo mass, but there is significant scatter in SMBH mass at fixed halo mass (the IQR of $M_{\rm BH}$ at $M_{200} = 10^{12.3}\,{\rm M}_\odot$ is 0.44 dex). This scatter correlates strongly with the residuals about the median $E_{\rm bind}-M_{200}$ relation, implying that halo binding energy is a more fundamental driver of SMBH mass than halo mass (\autoref{fig:mbh_m200}).

    \item We demonstrate that the scaling relation connecting SMBH mass and halo binding energy indeed exhibits less scatter than that connecting SMBH mass and halo mass (e.g. the IQR of $M_{\rm BH}$ at the median binding energy corresponding at $M_{200} = 10^{12.3}\,{\rm M}_\odot$ is 0.36 dex). Moreover we demonstrate that at fixed $E_{\rm bind}$, $M_{\rm BH}$ anti-correlates strongly with morphological diagnostics that quantify the `disciness' of the galaxy, such that more (less) disc-dominated galaxies host a less (more) massive SMBH (\autoref{fig:mbh_ebind}).
    
   \item We examine the evolution of the SMBH mass, the co-rotational stellar ($\kappa_{\rm co, \star}$) and gas ($\kappa_{\rm co,gas}$) kinetic energy fractions, and the ex-situ stellar mass fraction ($f_{\rm ex\text{-}situ}$) of the main progenitors of galaxies identified at the present-day as the most-discy and least-discy galaxies within the binding energy range that exhibits the greatest scatter in SMBH mass. We find that the SMBHs hosted by the least-discy sub-sample experiences a period of rapid growth that is broadly co-temporal with the decline of $\kappa_{\rm co, \star}$. In contrast, the SMBHs hosted by the most-discy sample experience no rapid growth phase. From inspection of the evolution of $f_{\rm ex\text{-}situ}$, we conclude that galaxies in the least-discy sample evolve secularly, with only $\sim 5$ percent of their present-day stellar mass acquired through mergers, compared to $\sim 25$ percent for the most-discy galaxies. We therefore infer that rapid SMBH growth is facilitated by mergers that disrupt the stellar disc and funnel gas to the galaxy centre (\autoref{fig:mbh_history}). We find that galaxy morphology can be influenced by tidal forces during ongoing mergers before the stellar mass is fully accreted, inducing SMBH growth through disruption of the galactic disc.
    
\end{itemize}

Our findings corroborate, using simulations that reproduce the key properties of the present-day galaxy population, the conclusion of \citet{Booth-Schaye2010} that SMBH mass is governed primarily by halo binding energy. Further, by showing that scatter about the $M_{\rm BH}-E_{\rm bind}$ relation anti-correlates with the `disciness' of galaxies, across the entire range of halo binding energies probed by the EAGLE Ref-L100N1504 simulation, we have generalised the findings of \citet{Davies2022,Davies2024}. Those authors used zoom simulations of an individual galaxy with controlled assembly histories, to show that galaxy discs provide rotational support that hinders the inward transport of gas towards the central SMBH, thus implying that disruptive mergers are, in general, needed to initiate the non-linear growth of SMBHs. Such growth is, in turn, necessary to enable SMBH self-regulation \citep[e.g.][]{McAlpine2018}, to expel a significant fraction of gas from the CGM \citep[e.g.][]{Davies2019,Oppenheimer2020}, and hence to quench galaxies \citep[e.g.][]{Appleby2020,Davies2020,Terrazas2020}. We thus provide a physical explanation for the observed influence of galaxy morphology on scatter about the scaling relation connecting the mass of central supermassive black holes to the stellar mass of their host galaxies.

Other cosmological simulations exhibit a dependence of SMBH mass on galaxy morphology (either at fixed stellar mass or velocity dispersion) that is, at least in some regimes, qualitatively similar to the relation exhibited by EAGLE \citep[e.g.][]{Li2020,Smethurst2024}. However, the current generation of such simulations yield diverse present-day $M_{\rm BH}-M_\star$ scaling relations \citep[e.g.][]{Habouzit2021}. It is well understood that the form of such relations (and the scatter about them) is sensitive to the detail of how feedback processes are implemented. We further caution that the simulations adopt a number of approximations concerning the seeding, growth (by both accretion and mergers) and dynamics of BHs, that severely restrict their predictive power \citep[e.g.][]{Tremmel2015,Bahe2021}. It will therefore be interesting to test whether future simulations that do not need to appeal to such restrictive approximations retain BH scaling relations that are sensitive to galaxy morphology.
\FloatBarrier

\section*{Acknowledgements}

We thank the referee, Kai Wang, for constructive comments that improved the study, and Victor Forouhar Moreno and Andrew Pontzen for insightful discussions. RJR acknowledges a PhD studentship at the LIV.INNO Centre for Doctoral Training “Innovation in Data Intensive Science”. This study was jointly supported by Liverpool John Moores University and Aspia Space Ltd. JJD acknowledges support from STFC grant ST/Y002482/1. RAC acknowledges support from STFC grants ST/Y002482/1 and ST/Y001907/1. This study made use of the Prospero high performance computing facility at Liverpool John Moores University.

\section*{Data Availability}

The EAGLE simulations have been released for public use. Galaxy and halo catalogues (accessible via SQL queries), and full particle data, can be accessed from the following url: \url{http://icc.dur.ac.uk/Eagle/database.php}. The reference article for the data release is \citet{McAlpine2016}, and the EAGLE team has posted a manual with more detailed guides explaining how to process the data \citep{EagleTeam2017}. The data used for the development of this article will be shared on reasonable request to the corresponding author.



\bibliographystyle{mnras}
\bibliography{References}





\bsp	
\label{lastpage}
\end{document}